\documentclass[twocolumn, amssymb, nobibnotes, aps, prd]{revtex4-2} 
\RequirePackage{xspace} % xspace
\usepackage{amsmath,amssymb}
\usepackage{mathrsfs}
\usepackage{bm}
\allowdisplaybreaks[4]
\usepackage{mathtools}
\usepackage{algpseudocode,algorithm}

\newcommand{\e}{\text{e}} % Nepia constant
\newcommand{\Tr}{\mathrm{Tr}\,}  %capital Trace
\newcommand{\av}[1]{\left\langle #1 \right\rangle} %Def of <average>
\newcommand{\Ket}[1]{\left| #1 \right\rangle} %Def of |i>
\renewcommand{\bar}[1]{\overline{#1}}%
\newcommand{\br}{\notag\\&\;\;\;\;\;\;}%change line in equation
\usepackage[dvipdfmx]{color}
\RequirePackage{ifpdf}
\ifpdf
  \usepackage[pdftex]{graphicx}
\else
  \usepackage[dvipdfmx]{graphicx}
\fi
\usepackage{cases}
\usepackage[utf8]{inputenc}
\usepackage[T1]{fontenc}
\usepackage[english]{babel}
\usepackage{ulem} %sout
\newcommand{\BetaVQE}{$\beta$-VQE\xspace}
\begin{document}
%
% Main-part !!
%
\title{Schwinger model at finite temperature and density with \BetaVQE}
\preprint{\today}
\author{Akio Tomiya}
\email[]{akio@yukawa.kyoto-u.ac.jp}
%\email[]{akio.tomiya@riken.jp}
\affiliation{Department of Information Technology, International Professional University of Technology, Osaka, 3-3-1, Umeda, Kita-Ku, Osaka, 530-0001, Japan}
%\affiliation{RIKEN/BNL Research center, Brookhaven National Laboratory, 
%Upton, NY, 11973, USA}
%\email[]{akio.tomiya@mail.ccnu.edu.cn}
%\affiliation{\small Key Laboratory of Quark \& Lepton Physics (MOE) and Institute of Particle Physics, Central China Normal University, Wuhan 430079, China}
%
\begin{abstract}
We investigate a quantum gauge theory at finite temperature and density using a variational algorithm for near-term quantum devices.
We adapt \BetaVQE to evaluate thermal and quantum expectation values and study the phase diagram for massless Schwinger model along with
the temperature and density.
By compering the exact variational free energy, we find the variational algorithm work for $T>0$ and $\mu>0$ for the Schwinger model.
No significant volume dependence of the variational free energy is observed in $\mu/g \in[0, 1.4]$.
We calculate the chiral condensate and take the continuum extrapolation.
As a result, we obtain qualitative picture of the phase diagram for massless Schwinger model.
%\textcolor{red}{No clear phase transition, in the theory}.
%By employing a hybrid variational algorithm $\beta$-VQE, we find \textcolor{red}{Hogehoge}.
%\begin{enumerate}
%\item \BetaVQE for the Schwinger model.
%\item continuum, large volume limit
%\item $T > 0$, $\mu \geq 0$
%\end{enumerate}
\end{abstract}
%\keywords{}

\maketitle


\section{Introduction}
Phase diagram of QCD  is one of the most attractive subject in particle physics and nuclear physics \cite{CABIBBO197567, Fukushima:2010bq, Guenther:2020jwe}.
At zero quark chemical potential, lattice QCD calculations with Markov-chain Monte-Carlo work well and we can access quantitative information, {\it e.g.} pseudo-critical temperature and cumulants \cite{HotQCD:2012fhj, Ding:2015ona, HotQCD:2018pds, Borsanyi:2018grb, Karsch:2019mbv, Goswami:2020yez}.
These values are tightly related to the heavy ion collision experiments and cosmological observations, and it is now more and more important to investigate QCD phase diagram to understand nature of our universe.
%The lattice calculations find the pseudo-critical temperature at the physical point and also the chiral limit and enables us to predict phenomena in experiments.
% Neural network is a powerful tool to investigate QCD phase diagram. At zero density $XXX$. 

Phase structure along with finite baryon density cannot be accessed the method above, due to the infamous sign problem, and a number of proposals are suggested (see \cite{Gattringer:2015nea, Ratti:2019tvj, Guenther:2020jwe, Alexandru:2020wrj} and references therein).
The complex Langevin algorithm has been applied not only matrix models and toy models, but also QCD in four dimension \cite{Aarts:2017vrv, Seiler:2017wvd, Berger:2019odf, Scherzer:2020kiu}. 
Lefschetz thimble is a similar approach, but it is an exact algorithm \cite{Witten:2010cx, Mori:2017zyl, Schmidt:2017gvu, Fujisawa:2021hxh, Fukuma:2022yhy}. 
Methods with change of variable is also investigated \cite{ Mori:2017nwj, Alexandru:2017czx, Kashiwa:2018vxr, Lawrence:2021izu, Detmold:2021ulb, Kanwar:2021tkd, Namekawa:2021nzu}.

Algorithms for digital quantum computers for lattice QCD are highly demanded to overcome the sign problem \cite{Lawrence:2020irw}.
Quantum algorithms can be regarded as ways to realize a unitary operation on quantum state based on unitary operations on qubits.
Quantum algorithms for quantum field theory has been applied for the real time \cite{Martinez:2016yna, Huffman:2021neh} and $\theta$ vacuum \cite{Chakraborty:2020uhf}. 

\begin{figure}[ht]
\begin{center}
\includegraphics[width=0.5\textwidth]{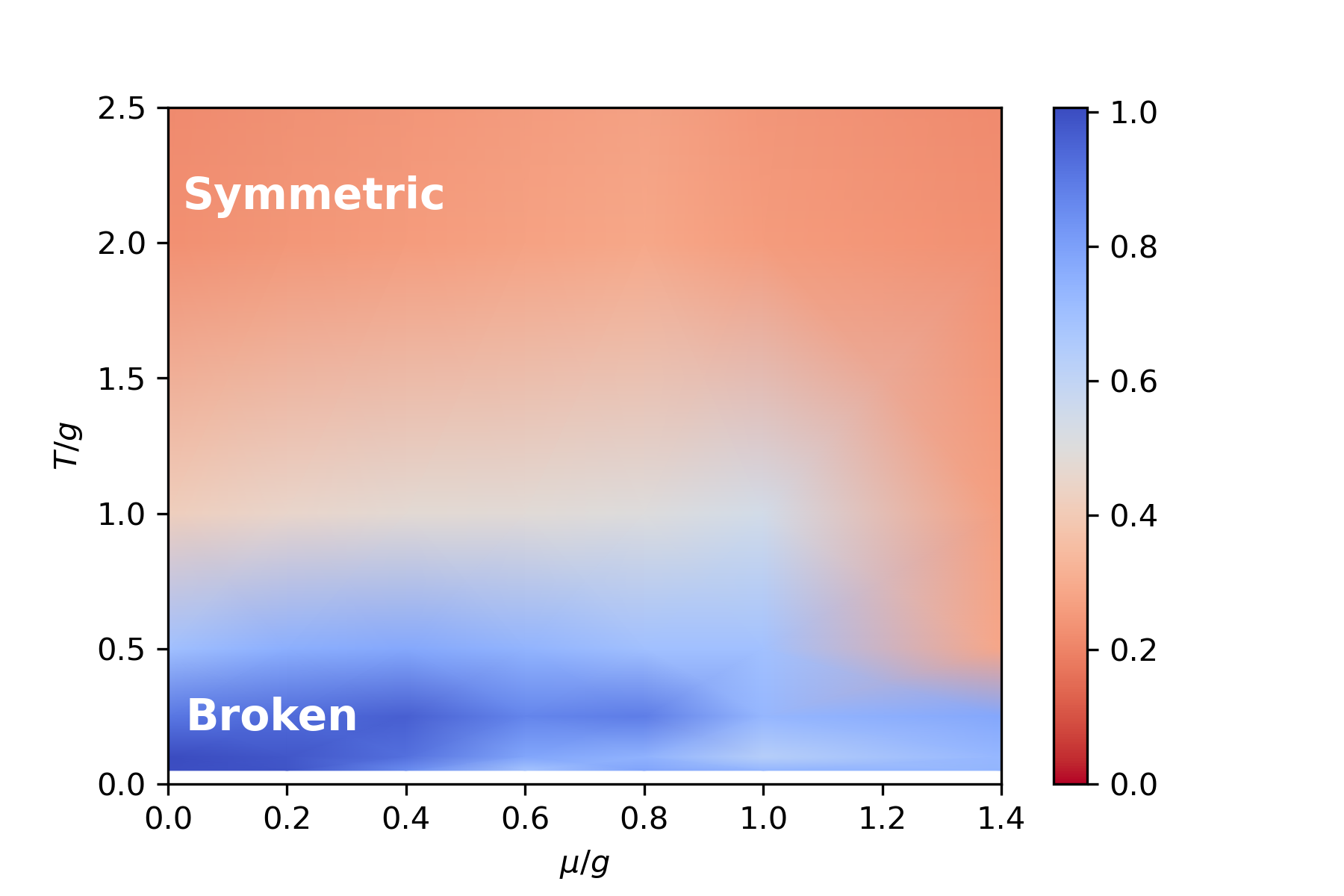}
\end{center}
\caption{
Density plot of the chiral condensate from \BetaVQE along with $T/g$ -- $\mu/g$.
The central values in Fig. \ref{fig:PbP_beta_nx10} are interpolated and shown (see the main text in detail).
%The results for the chiral condensate are normalized with the exact value in the continuum at $m=0$ with $T=0$
This density plot can be seen as the phase diagram. 
\label{fig:phase-diagram}}
\end{figure}

Noisy Intermediate-Scale Quantum Computers (NISQ) \cite{Bharti:2021zez}
have limited capability than ones with error correction, but still it is useful to investigate quantum field theory since it enables us to investigate physics at parameters with sign problem \cite{Yamamoto:2021vxp}.
While it has been investigated purely quantum algorithm \cite{motta2020determining}, classical-quantum hybrid variational algorithms is considerably suitable for NISQ devices
\cite{Liu_2019, Ciavarella:2021lel} rather than exact algorithms.

Quantum field theory at finite temperature on digital quantum computers is not straightforward because the quantum circuit only realizes the unitary evolution.
Conventional  thermodynamics uses mixed states, and it cannot be realized using unitary operations only. 
Naively, we can combine the unitary operations and projective measurements to realize the mixed states, but we loose signal exponentially in the volume.
On the other hand, we can eigenstate thermalization hypothesis \cite{Deutsch_2018} or the thermal pure quantum state argument \cite{Sugiura:2011hm} which can be realized in the unitary operations/quantum circuit but it is not widely investigated in this context.

A classical quantum hybrid variational algorithm \BetaVQE \cite{liu2021solving} for NISQ devices enables us to calculate thermal-quantum average without the sign problem and exponential signal lost.
It uses a parametrized probability distribution to emulate thermal part of the density matrix, and quantum device or emulator to calculate matrix elements. 
Moreover, for small system, we can evaluate the quality of the variational approximation on a classical computer because it becomes exact if
the variational free energy is zero \cite{liu2021solving}.
It has been applied on the transverse filed Ising model and it agree with the exact result.

The  Schwinger model \cite{Schwinger:1962tp, Pak:1977an, Johnson:1963vz}
is a suitable testbed to examine a new algorithm, which is two dimensional quantum abelian gauge theory with fermionic matters which has common features to QCD in four dimension.
It is exactly solvable at $m=0$ \cite{Schwinger:1962tp, Sachs:1991en}.
For zero temperature in the path integral formalism, lattice version of the model is investigate to explore new algorithms with machine learning \cite{Finkenrath:2022ogg, Albergo:2022qfi}.
For hamiltonian formalism, a tensor network has been used to investigate finite temperature or finite chemical potential
\cite{Banuls:2013jaa, Banuls:2016lkq, Funcke:2019zna, Butt:2019uul, Fukuma:2021cni}.
%[Saito, Jansen, +?].
Recently it has been employed to examine feasibility of application of lattice field theory  on quantum computers
\cite{Martinez:2016yna, Chakraborty:2020uhf, Shaw:2020udc, Huffman:2021neh, Yamamoto:2021vxp}.

In this work, we investigate phase structure of the massless Schwinger model for finite temperature and density toward to investigate QCD phase diagram using a classical-quantum hybrid algorithm.
More precisely, we evaluate $T$--$\mu$ dependence of the chiral condensate with \BetaVQE on a classical computer.
We measure the variational free energy to evaluate quality of variational thermal states. 
%By using the variational states and classical distribution, we calculate the chiral condensate of the Schwinger model at finite temperature and density using classical-quantum hybrid algorithm, and w
We evaluate the large volume limit with data from two large volume and take the continuum limit.
As a result, we obtain a phase diagram in  Fig. \ref{fig:phase-diagram}.

This paper is organized as follows.
In next section, we introduce the Schwinger model and its the spin representation. In addition, we review \BetaVQE briefly.
Next, we show the results. We discuss the variational free energy, which quantify the quality of the approach.
In next section, we evaluate the large volume limit and the continuum limit.
In next section, we show the results for the chiral condensate as a function of the coupling constant and the chemical potential.
Finally, we summarize this work.

\section{Model and algorithm}
\subsection{Schwinger model and spin representation}
Here we explain one flavor Schwinger model briefly, and please refer \cite{Chakraborty:2020uhf} for details.
The Lagrangian of  the Schwinger model with $\theta = 0$ is given by
\begin{align}
\mathcal{L}=-\frac{1}{4} F_{\mu \nu} F^{\mu \nu}+\mathrm{i} \bar{\psi} \gamma^{\mu}\left(\partial_{\mu} +\mathrm{i} g A_{\mu}\right) \psi-m \bar{\psi} \psi,
\end{align}
where $g$ is the dimension-full coupling constant.
To write down the hamiltonian, we have to fix the gauge
and we choose the temporal gauge $A_0(x)=0$.
The hamiltonian is,
\begin{align}
H
&=\int d x\Big[-\mathrm{i} \bar{\psi} \gamma^{1}\left(\partial_{1}+\mathrm{i} g A_{1}\right) \psi
+m \bar{\psi} \psi+\frac{1}{2} \Pi^{2} \Big],
\end{align}
where $\Pi \equiv \dot{A}^{1}$. % and ${L_x}$ is the spatial extent and we choose the boundary condition later.
In addition to the Hamiltonian, we need Gauss' law to make the Hilbert space to keep be physical,
\begin{align}
0=-\partial_{1} \Pi-g \bar{\psi} \gamma^{0} \psi.
\end{align}

We are interested in temperature and chemical potential dependence of thermal and quantum expectation values of an operator $O$,
\begin{align}
\av{O} = \frac{1}{Z}\Tr[O \e^{-\frac{1}{T} H}] = \Tr[O \rho]
\end{align}
where $\rho =\e^{-\frac{1}{T} H}/Z$ and $Z$ is a normalization constant and $T$ is the temperature.
$\Tr[\cdots]$ means functional trace over the Hilbert space.

The Schwinger model enjoys non-zero chiral condensate at the ground state and also finite temperature $T$ \cite{Sachs:1991en},
\begin{align}
\av{\bar \psi \psi}(T) = \Sigma_0 \e^{I(m_g/T) }
\end{align}
where 
$ \Sigma_0= m_g \e^\gamma/(2\pi)$ is the chiral condensate at zero-temperature and $m_g = g/\sqrt{\pi}$.
The function $I(x)$ is,
\begin{align}
I(x) = \int_0^\infty \frac{1}{1- e^{x\cosh t } } dt.
\end{align}
This has been calculated using the tensor network.

We descritize the system with the staggered formalism,
\begin{align}
H
&=-\mathrm{i} \sum_{n=1}^{N_x-1}w\left[\chi_{n}^{\dagger} e^{\mathrm{i} \phi_{n}} \chi_{n+1}-\text { h.c. }\right] 
\br
+
\sum_{n=1}^{N_x} m (-1)^{n} \chi_{n}^{\dagger} \chi_{n}+J \sum_{n=1}^{N_x-1} L_{n}^{2}
\end{align}
where 
$w=1 /(2 a) $ $J=g^{2} a / 2$, and and $N_x$ is dimensionless spatial extent of the staggered fermions.
$L_{n}-L_{n-1}=\chi_{n}^{\dagger} \chi_{n}-({1-(-1)^{n}})/{2}$ is descritized Gauss' law and $L_1$ chosen by the boundary condition.
By imposing the open boundary condition and apply a gauge transformation, we can eliminate the gauge field from the Hamiltonian and as the drawback, 
we obtain a non-local interaction term.

We can include the number operator $\chi_{n}^{\dagger} \chi_{n}$ with the chemical potential $\mu$ to the hamiltonian as \cite{Banuls:2016gid}, 
\begin{align}
%H
%&=-\mathrm{i} \sum_{n=1}^{N_x-1}w\left[\chi_{n}^{\dagger} e^{\mathrm{i} \phi_{n}} \chi_{n+1}-\text { h.c. }\right] 
%\br+
\sum_{n=1}^{N_x }  (m (-1)^{n}  )  \chi_{n}^{\dagger} \chi_{n}
\Rightarrow
\sum_{n=1}^{N_x }  (m (-1)^{n} + \mu )  \chi_{n}^{\dagger} \chi_{n}.
%+J \sum_{n=1}^{N_x-1} L_{n}^{2}
\end{align}

In order to calculate on digital quantum computers, operators in the hamiltonian must be written spin operators.
The Jordan-Wignar transformation makes the spin representation of fermions,
\begin{align}
\chi_{n}=\left(\prod_{  1 \leq l <n}-\mathrm{i} Z_{l}\right) \frac{X_{n}-\mathrm{i} Y_{n}}{2},
\end{align}
where $X_n$, $Y_n$ and $Z_n$ are Pauli matrices for $x, y, z$ direction acting on $n$-th qubit.
This transformation absorbs difference of statistical property between Pauli matrices and fermion fields.

%After the transformation we get the Hamiltonian for the Schwinger model in spin operator representation,
%\begin{align}
%H = H_{ZZ} + H_\pm + H_Z.
%\end{align}
%where
%\begin{align}
%H_{Z Z} &=\frac{J}{2} \sum_{n=2}^{N-1} \sum_{1 \leq k<l \leq n} Z_{k} Z_{l},\\
%H_{\pm} &=\frac{1}{2} \sum_{n=1}^{N-1}w\left[X_{n} X_{n+1}+Y_{n} Y_{n+1}\right],\\
%H_{Z} &=\frac{1}{2} \sum_{n=1}^{N} (m(-1)^{n} +\mu )Z_{n}-\frac{J}{2} \sum_{n=1}^{N-1}(n \bmod 2) \sum_{l=1}^{n} Z_{l}
%\end{align}
After the transformation, we get the Hamiltonian for the Schwinger model in spin operator representation,
\begin{align}
\hat{H} = H/g= \hat{H}_{ZZ} + \hat{H}_\pm + \hat{H}_Z.
\end{align}
where
\begin{align}
&\hat{H}_{Z Z} =\frac{g}{8w} \sum_{n=2}^{N_x-1} \sum_{1 \leq k<l \leq n} Z_{k} Z_{l},\\
&\hat{H}_{\pm} =\frac{1}{2} \sum_{n=1}^{N_x-1}\frac{w}{g}\left[X_{n} X_{n+1}+Y_{n} Y_{n+1}\right],\\
&\hat{H}_{Z} =\frac{1}{2} \sum_{n=1}^{N_x} \left(\frac{m}{g}(-1)^{n}  +\frac{\mu}{g} \right)Z_{n} \br 
-\frac{g}{8w} \sum_{n=1}^{N_x-1}(n \bmod 2) \sum_{l=1}^{n} Z_{l}
\end{align}
and $\hat{\cdot}$ indicates dimensionless quantity. 
The exponent of Boltzmann weight is now $H/T = \hat{H} g/T$, and $g/T$ is dimensionless inverse temperature,
which conventionally referred as $\beta$ but we do not use this notation to avoid confusion with $1/g^2$.
%We can read off the dimensionality of the Hilbert space and it is $2^{N_x}$.

In this study, we do not use the adiabatic state preparation \cite{Chakraborty:2020uhf}, instead, we use a classical-quantum hybrid algorithm \BetaVQE,
to evaluate thermal and quantum expectation values.
We focus on $m=0$ in the numerical calculations, which does not have additive divergence of the chiral condensate.

\subsection{\BetaVQE: Neural network parametrized ansatz}
In this work, we adapt \BetaVQE to evaluate quantum and thermal expectation value \cite{liu2021solving}.
Here we briefly review \BetaVQE based on the original paper.
VQE stands for variational quantum eigen-solver, which enables us to evaluate quantum expectation values with a parametrized shallow quantum circuit. 
The shallow quantum circuit is parametrized using rotation gates, which rotates a qubit in SU(2) space.
The initial state is taken to classically tractable product state and a shallow quantum circuit produces entanglement.
The parameters in the circuit is tuned using classical process to minimize a loss function including quantum expectation value
which is evaluated quantum computer or quantum circuit emulator.
We do not pursue exactness since the variational method gives precise results based on the variational principle but is not an exact method in practice, and results must be qualitative. 

We consider a $N$-qubit system represented by a bit string $x = x_1 x_2 \cdots x_N$ where $x_i \in \{0,1\}$ ($i=1,\cdots,N$).
In particular we employ a product state,
\begin{align}
\Ket{x} =  \Ket{x_1}\otimes \Ket{x_2}\otimes \cdots \otimes \Ket{x_N} .
\end{align}
We call this as the classical product state, which is easy to prepare.
Variational approaches use a parametrized state $U_\theta\Ket{x} $, where $\theta$ is a set of parameters for a quantum state
and $U_\theta$ a unitary evolution like rotation gates in quantum circuits.
In practice, we use SU(4) parametrization for $U_\theta$ as in the original work \cite{liu2021solving}.

We parametrize the density matrix as, % acts on a state $\Ket{x}$ as 
\begin{align}
\tilde{\rho}_\Theta = \sum_x p_{\phi}(x) U_{\theta}|x\rangle\langle x| U_{\theta}^{\dagger}, \label{eq:parametrized_state}
\end{align}
where  $\Theta = \theta {}^\cup \phi$ and $p_\phi$ is a parametrized classical distribution.
%In physicists language, this $p_\phi$ is the Boltzmann distribution for an effective model.
This density matrix satisfies the normalization condition $\Tr[\tilde{\rho}_\Theta ] = \sum_x p_{\phi}(x) = 1$.
In the application, we employ the autoregressive model \cite{autogressive2015}, which is a neural network.
As a unit of the autoregressive model, we utilize an autoencoder, which has single hidden layer of 500 hidden neurons with rectified linear unit activation function. 
This is also the same setup from the original work for \BetaVQE \cite{liu2021solving}.
They show that, \BetaVQE works well for two dimensional transverse field Ising model.

Our target system is written as $\rho =\e^{-\frac{1}{T} H}/Z$ and we minimize following loss function,
\begin{align}
\mathcal{L}(\Theta) 
= \bar{D}( \tilde{\rho}_\Theta || \rho)
=\operatorname{Tr} \left(\tilde{\rho}_\Theta \ln \left( \frac{\tilde{\rho}_\Theta}{ \rho } \right) \right),%\\
%&=\operatorname{Tr} \left(\tilde{\rho}_\Theta \ln \left( \frac{\tilde{\rho}_\Theta}{ \e^{-\frac{1}{T} H}/Z} \right) \right),%\\
\end{align}
by tuning parameters $\Theta$. 
This variational argument is developed in the Gibbs-Delbruck-Moliere variational principle of quantum statistical mechanics \ref{huber1968variational}
We remark that, this loss function can be seen as,
quantum version of (reversed) Kullback Leibler divergence (an application for lattice field theory with the classical version can be found in 
\cite{Kanwar:2020xzo,Albergo:2019eim,Boyda:2020hsi}
).
The Kullback Leibler divergence for classical probability distributions $p, q$ is given by,
\begin{align}
{D}( p || q) = \int d x\;  p(x) \ln \frac{p(x)}{q(x)},
\end{align}
which is referred as the relative entropy in physics context.
As same as the classical Kullback Leibler divergence ${D}( p || q) $,
this $\bar{D}( \tilde{\rho}_\Theta || \rho)$ is zero if and only if $\tilde{\rho}_\Theta  = \rho$ \cite{liu2021solving}. 
This means that, we can quantify the quality of variational solution through $\mathcal{L}(\Theta)$.

%\begin{align}
%\mathcal{L}(\Theta) %&=\operatorname{Tr}(\tilde{\rho}_\Theta \ln \tilde{\rho}_\Theta)-\operatorname{Tr}(\tilde{\rho}_\Theta  \ln \e^{-\frac{1}{T} H}) + \ln Z,\\
%=\operatorname{Tr}(\tilde{\rho}_\Theta \ln \tilde{\rho}_\Theta)+\frac{1}{T} \operatorname{Tr}(\tilde{\rho}_\Theta H) + \ln Z, %\\
%%&=\operatorname{Tr}(\tilde{\rho}_\Theta \ln \tilde{\rho}_\Theta)-\operatorname{Tr}(\tilde{\rho}_\Theta \log \e^{-\frac{1}{T} H}) +  \Tr[\tilde{\rho}_\Theta \ln Z ], \\
%\end{align}

%To use \BetaVQE o
On the real quantum devices as intended in the original paper,  we cannot calculate the partition function $Z$, or equivalently the free energy $-\ln Z$.
In that case, we have to use reversed and shifted loss function,
\begin{align}
\mathcal{L}(\Theta) - \ln Z 
%&= \operatorname{Tr} \left(\tilde{\rho}_\Theta \ln \left( \frac{\tilde{\rho}_\Theta}{ \rho } \right) \right) - \ln Z ,\\
&=\operatorname{Tr}(\tilde{\rho}_\Theta \ln \tilde{\rho}_\Theta)+\frac{1}{T} \operatorname{Tr}(\tilde{\rho}_\Theta H) \label{eq:shift_var_energy}.
\end{align}
%We note that, this still does not contain bias, namely if we find a parameter $\Theta$ which globally minimizes ${\mathcal{L}}(\Theta)$, it is exact.
We remark that no bias property is still held. Namely, this loss function is minimized if and only if $\tilde{\rho}_\Theta  = \rho$.
In addition, the constant shift does not affect on the derivative and training and the right hand side is calculable because it is free from the normalization constant.
However in this case, we cannot see that the parameter $\Theta$ is optimal or not.
In this pilot study, we calculate $\mathcal{L}(\Theta)  = \hat{D}( \tilde{\rho}_\Theta || \rho)$, not the shifted one, using 
state vector to examine the algorithm and
observe the phase structure.

Since it can produce training samples by itself, we rely on the self-training paradigm for the training 
We sample from the model distribution as, in the training process,
\begin{align}
{\mathcal{L}}(\Theta) = \mathbb{E}_{x \sim p_{\phi}(x)} \left[\ln p_{\phi}(x)+\frac{1}{T}\left\langle x\left|U_{\theta}^{\dagger} H U_{\theta}\right| x\right\rangle\right] +{\rm const}.
\end{align}
Optimization of the loss function ${\mathcal{L}}(\Theta)$ is performed by a stochastic gradient optimizer ({\it e.g.}, adam \cite{adamopt}).
We have two kinds of derivative. One is a derivative with respect to parameters in the quantum circuit,
\begin{align}
\nabla_{\theta} {\mathcal{L}}(\Theta) = \frac{1}{T} \mathbb{E}_{x \sim p_{\phi}(x)}\left[\nabla_{\theta}\left\langle x\left|U_{\theta}^{\dagger} H U_{\theta}\right| x\right\rangle\right].
\end{align}
This derivative can be evaluated by the shift rule \cite{ shiftrule1, Banchi_2021}.

The other is a derivative with respect to parameters in the classical distribution,
\begin{align}
\nabla_{\phi} {\mathcal{L}}(\Theta) =\mathbb{E}_{x \sim p_{\phi}(x)}\left[(f(x)-b) \nabla_{\phi} \ln p_{\phi}(x)\right],
\end{align}
where 
\begin{align}
f(x)=\ln p_{\phi}(x)+
\frac{1}{T}\left\langle x\left|U_{\theta}^{\dagger} H U_{\theta}\right| x\right\rangle
\end{align}
and $b=\mathbb{E}_{x \sim p_{\phi}}[f(x)]$.
This term $b$ is not essentially needed  but it helps to reduce the variance of the optimizer \cite{liu2021solving}.

After training, one can sample a batch of input states $\Ket{x}$ and
treat them as approximations of the eigenstates of the system.
Since the unitary circuit preserves orthogonality of the input
states, the sampled quantum states span a low energy subspace
of the Hamiltonian.

First we stochastically and iteratively find $\Theta^* =  \underset\Theta{\rm argmin} \mathcal{L}(\Theta)$.
By using $\Theta^*$ and \eqref{eq:parametrized_state}, we can evaluate the expectation values with sampling $x$ from $p_\phi$.
We evaluate spatially averaged chiral condensate \cite{Chakraborty:2020uhf},
\begin{align}
\av{\bar \psi \psi}_{\Theta^*} \big/g = \frac{2 w}{N_x}\sum_n \Tr[ \bar \psi \psi (n) \tilde{\rho}_{\Theta^*}  ].
\end{align}
%to stabilize the statistical property.
Note that, states, in particular for finite density, might not have translational invariance \cite{Metlitski:2006id, Maedan:2009yi, Narayanan:2012qf}
but we use this operator as an indicator.

To evaluate the quantum expectation value, we utilize calculations with state vectors on a classical computer to avoid uncertainty of circuit operations.
Use of classical computer limits the maximum number of sites in practice but it enables us to evaluate exact free energy.
In addition, the replacement is straightforward except for error evaluation of the variational method \cite{liu2021solving}.

%$x$ is a bit string and a probability distribution is written as an energy based model,
%\begin{align}
%p_\phi(x) = f_\phi(x)
%\end{align}

\section{Results}

We perform \BetaVQE calculations for following parameters:
$N_x \in  \{ 4, 6, 8, 10\}$ (dimensionless spatial extent of the spin model),
$w/g \in  \{0.5, 0.45, 0.4, 0.35\} $ (lattice spacing),
$g/T  \in \{ 0.1, 0.5, 1.0, 2.0, 4.0, 10.0, 20.0 \}$  (inverse temperature),
$\mu/g = \{ 0.0, 0.2, 0.4 , 0.6, 1.0, 1.4\}$ (chemical potential). % , 0.8
Our temperature range is $T/g \in [0.05, 10]$.
As we mentioned, we store a state vector in this study % to avoid systematics from quantum devices
to get quantum expectation value in each step in the training,
and we cannot take the volume more than $N_x = 10$ in practice. %, that has $2^{2\times 10}$ dimensional Hilbert space.

The number of training is 500 for each $(N_x$, $w/g$, $g/T$, $\mu/g)$
and we employ the adam optimizer to minimize the loss function.
During the  training, we monitor the variational free energy to quantify the quality of the variational density matrix.
The number of sample for thermal average is taken to 5000.

In the analysis of the chiral condensate, %, we first evaluate the large volume limit ($w/N_x \propto 1/L_x \to 0$ under fixed $w$),
%and next 
we take the continuum extrapolation ($g/(2w) = ag \to 0$) for our largest lattice data.
%As we will comment, we discard data from $g/T = 20.0$ in the analysis for continuum physics.
As we will comment, we do not include data from $N_x = 4, 6$ in the analysis for continuum physics to avoid finite volume effects.
Detailed analysis with physically larger lattice using other method is left for another publication.

The calculations are carried out using an open source package \BetaVQE.jl \cite{betaVQEjl},
 which is based on Yao.jl \cite{Luo2020yaojlextensible}, a package for quantum calculations, and 
Zygote.jl \cite{Zygotejl}, a package for automatic differentiation in neural networks, in Julia language \cite{julialang}.

\subsection{Variational free energy in \BetaVQE}
Here we discus the variational free energy in \BetaVQE, $\mathcal{L}(\Theta)$.
%and  the number of epochs is taken to 500 for all parameters.

Tab \ref{tab:variational_free_energy} is the results of the variational free energy for our smallest volume $N_x = 4$ and largest volume $N_x=10$
with the cutoff $w=0.5$ and $0.35$ and the temperature $g/T = 0.1$ to $20.0$.
The variational free energy defined in \eqref{eq:shift_var_energy} in the end of the training at epoch 500 are listed.
A column Diff is $ (\mathcal{L}/(-\ln Z)) \times 100$, deviation from the exact free energy.
As we mentioned above, $\mathcal L$ is zero if and only if the variational approximation is exact.
Except for $g/T = 20, \mu/g=0$, $N_x$ = 4,  difference of loss function to the exact free energy density takes 0 in $O(1)$ \%.
In our parameter regime, we do not observe strong dependence on the chemical potential.
Rather, we observe strong dependence on the temperature, as expected.

\begin{figure}[h]
\begin{center}
\includegraphics[width=0.45\textwidth]{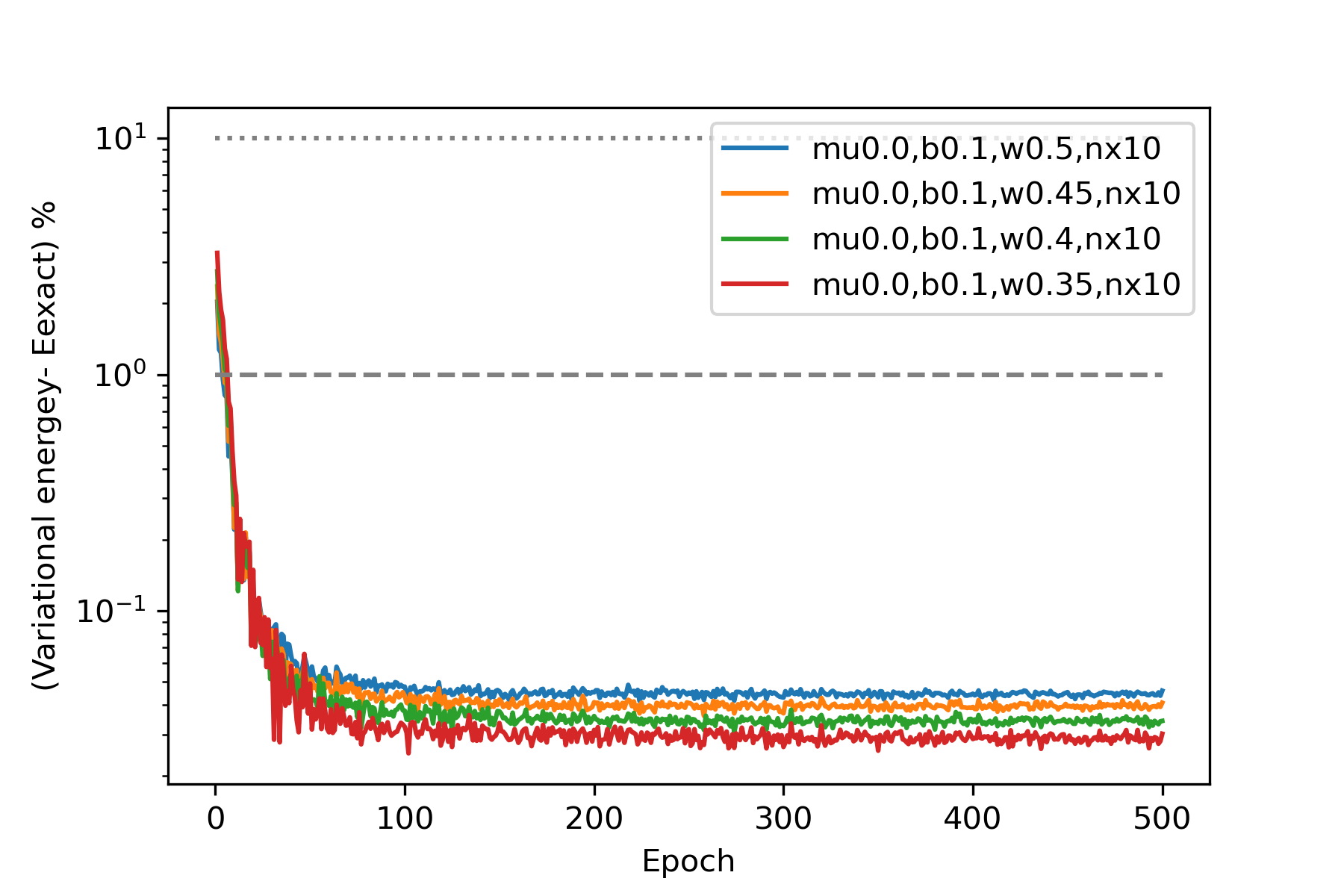}
\end{center}
\caption{
Deviation of the variational energy from the exact one as a function epoch for $N_x = 10$, $g/T = 0.1$ and $\mu/g=0.0$.
Blue, orange, green, red lines correspond to $w = 1/(2a) = 0.5, 0.45, 0.4 $ and $ 0.35$,\ respectively.
\label{fig:variational_energy_beta0.1mu0}}
\end{figure}

\begin{figure}[h]
\begin{center}
\includegraphics[width=0.45\textwidth]{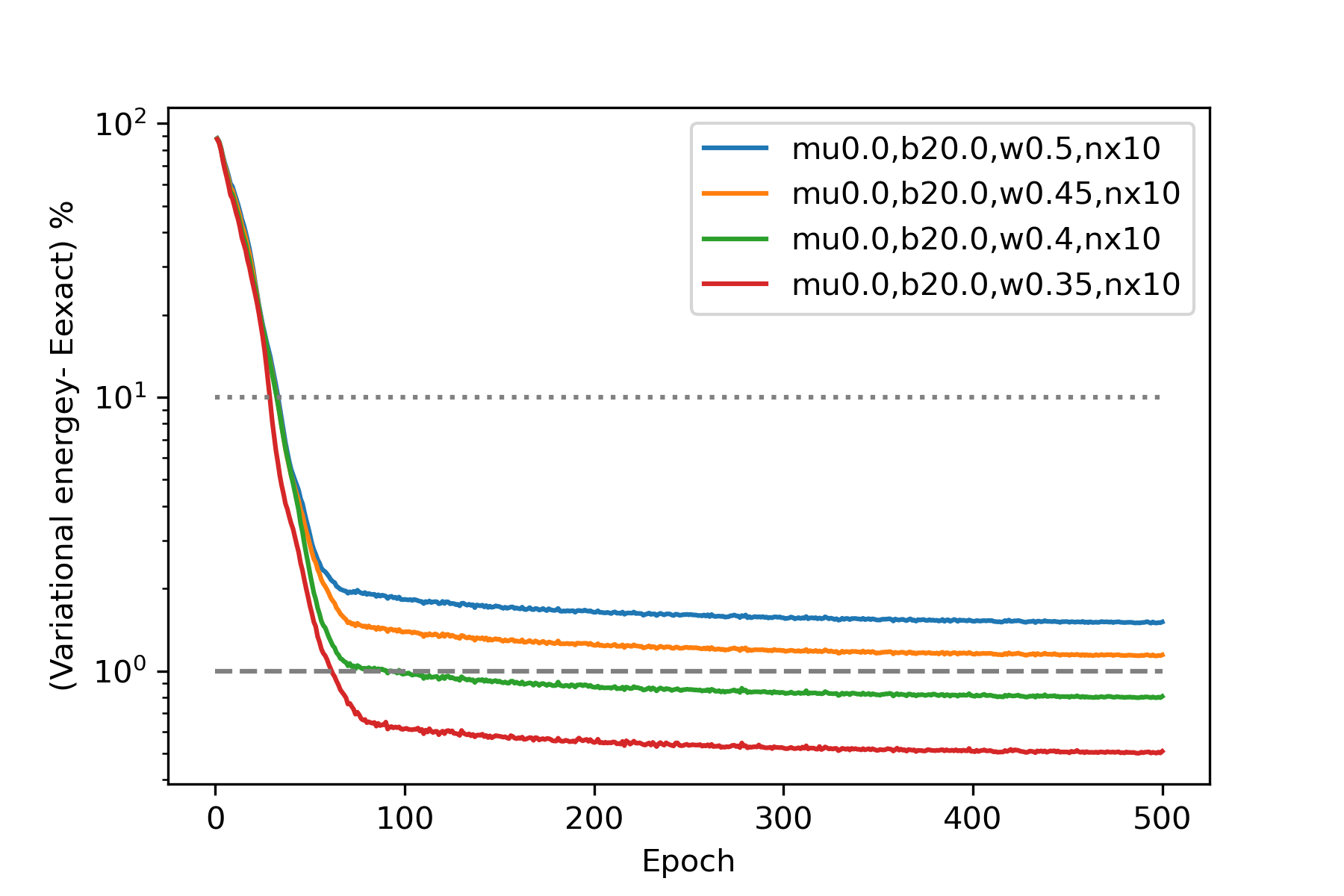}
\end{center}
\caption{
Same figure of 
Fig. \ref{fig:variational_energy_beta0.1mu0}
but for $g/T = 20$ and $\mu/g=0.0$.
\label{fig:variational_energy_beta20mu0}}
\end{figure}

\begin{figure}[h]
\begin{center}
\includegraphics[width=0.45\textwidth]{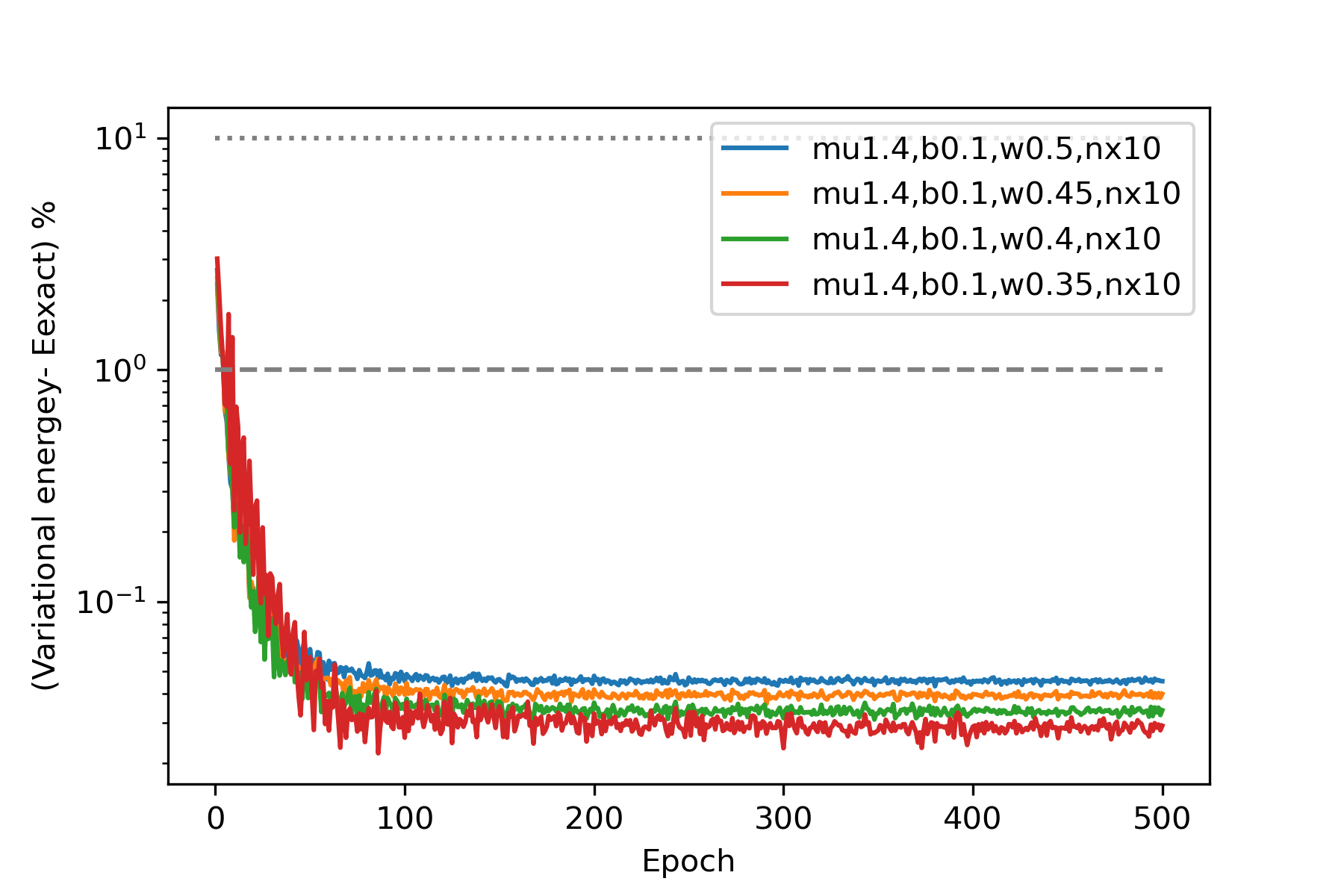}
\end{center}
\caption{
Same figure of 
Fig. \ref{fig:variational_energy_beta0.1mu0}
but for $g/T = 0.1$ and $\mu/g=1.4$.
\label{fig:variational_energy_beta0.1mu1.4}}
\end{figure}

\begin{figure}[h]
\begin{center}
\includegraphics[width=0.45\textwidth]{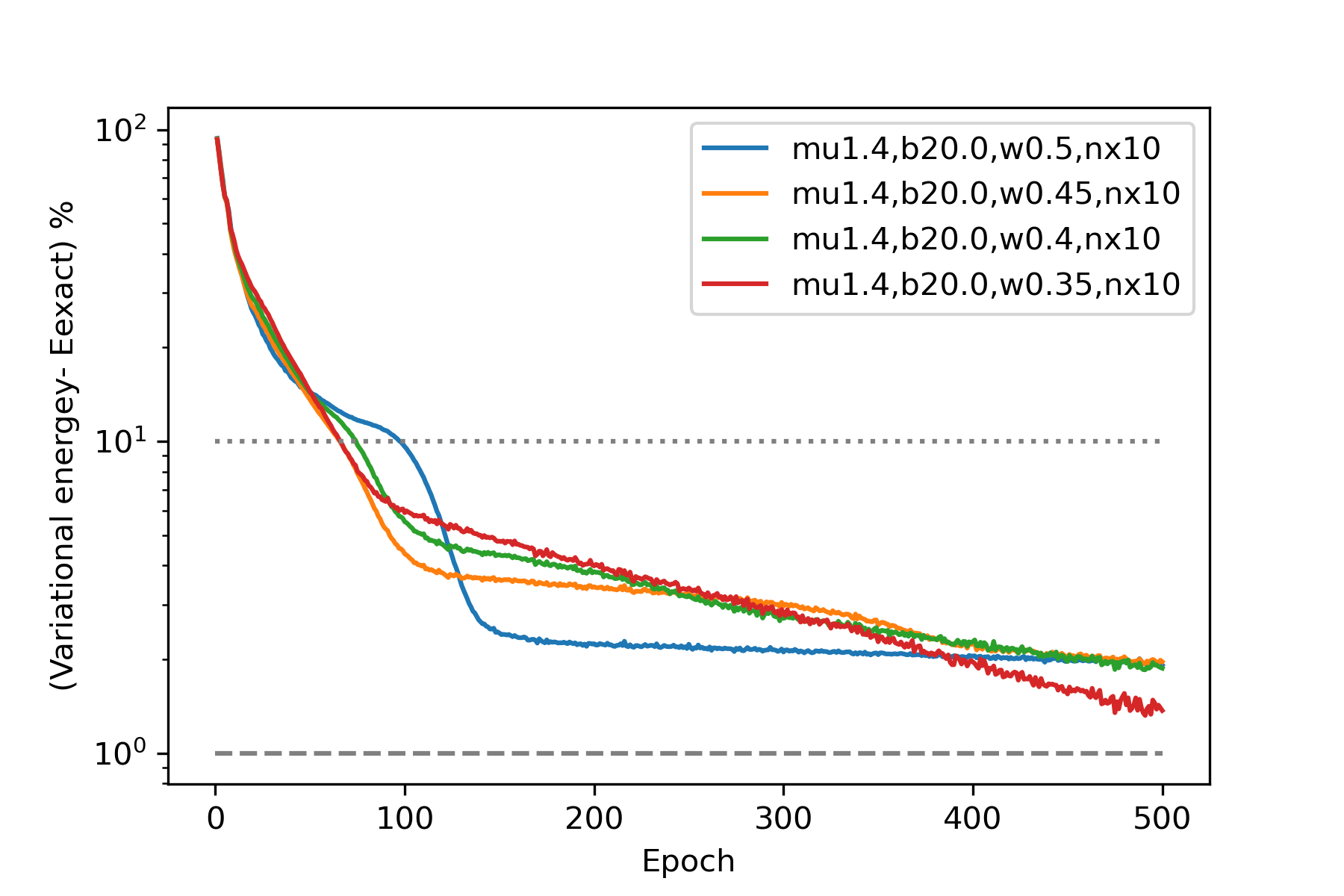}
\end{center}
\caption{
Same figure of 
Fig. \ref{fig:variational_energy_beta0.1mu0}
but for $g/T = 20$ and $\mu/g=1.4$.
\label{fig:variational_energy_beta20mu1.4}}
\end{figure}

We  plot deviation of the variational energy from the exact one as a function epoch in 
Fig. \ref{fig:variational_energy_beta0.1mu0}, 
Fig. \ref{fig:variational_energy_beta20mu0}, 
Fig. \ref{fig:variational_energy_beta0.1mu1.4}
and Fig. \ref {fig:variational_energy_beta20mu1.4} for  $N_x =10$, $g/T =0.1, 20$ and $\mu/g=0.0, 1.4$.
In each plot, blue, orange, green, red lines correspond to $w = 1/(2a) = 0.5, 0.45, 0.4 $ and $ 0.35$,\ respectively.
One can see that, the deviation is $O(1)$ \% after 500 epoch.
This also indicates that, the error from the variational method mainly depends on the temperature rather than the chemical potential in our parameter regime.

\begin{table}[htb]
%\begin{tabular}{c|c|c|c||c|c|c} 
%$\mu/g $ & $g/T$ & $N_x$ & $w/g$ & Diff (\%) \\ \hline \hline
%0.0 & 0.1 & 8 & 0.7 & 0.0477 \\
%0.0 & 0.1 & 8 & 0.55 & 0.0386 \\
%0.0 & 0.1 & 10 & 0.7 & 0.0646 \\
%0.0 & 0.1 & 10 & 0.55 & 0.0499 \\ \hline
%0.0 & 0.5 & 8 & 0.7 & 0.916 \\
%0.0 & 0.5 & 8 & 0.55 & 0.671 \\
%0.0 & 0.5 & 10 & 0.7 & 1.29 \\
%0.0 & 0.5 & 10 & 0.55 & 0.773 \\ \hline
%0.0 & 10.0 & 8 & 0.7 & 0.278 \\
%0.0 & 10.0 & 8 & 0.55 & 0.103 \\
%0.0 & 10.0 & 10 & 0.7 & 2.75 \\
%0.0 & 10.0 & 10 & 0.55 & 1.84 \\ \hline
%0.0 & 20.0 & 8 & 0.7 & 0.485 \\
%0.0 & 20.0 & 8 & 0.55 & 0.272 \\
%0.0 & 20.0 & 10 & 0.7 & 2.8 \\
%0.0 & 20.0 & 10 & 0.55 & 1.88 \\ \hline\hline
%%
%1.0 & 0.1 & 8 & 0.7 & 0.0564 \\
%1.0 & 0.1 & 8 & 0.55 & 0.0441 \\
%1.0 & 0.1 & 10 & 0.7 & 0.0669 \\
%1.0 & 0.1 & 10 & 0.55 & 0.0505 \\ \hline
%1.0 & 0.5 & 8 & 0.7 & 0.907 \\
%1.0 & 0.5 & 8 & 0.55 & 0.574 \\
%1.0 & 0.5 & 10 & 0.7 & 1.22 \\
%1.0 & 0.5 & 10 & 0.55 & 0.812 \\ \hline
%1.0 & 10.0 & 8 & 0.7 & 0.842 \\
%1.0 & 10.0 & 8 & 0.55 & 0.737 \\
%1.0 & 10.0 & 10 & 0.7 & 3.55 \\
%1.0 & 10.0 & 10 & 0.55 & 2.48 \\ \hline
%1.0 & 20.0 & 8 & 0.7 & 1.0 \\
%1.0 & 20.0 & 8 & 0.55 & 0.92 \\
%1.0 & 20.0 & 10 & 0.7 & 4.26 \\
%1.0 & 20.0 & 10 & 0.55 & 2.9 \\
\begin{tabular}{c|c|c|c||c|c|c} 
$\mu/g $ & $g/T$ & $N_x$ & $w/g$ & $ \mathcal{L}-\ln Z $ &  $-\ln Z$ & Diff (\%) \\ \hline \hline
0.0 & 0.1 & 4 & 0.5 & -27.779 & -27.781 & 0.00804 \\
0.0 & 0.1 & 4 & 0.35 & -27.807 & -27.808 & 0.005 \\
0.0 & 0.1 & 10 & 0.5 & -70.686 & -70.718 & 0.0459 \\
0.0 & 0.1 & 10 & 0.35 & -71.744 & -71.765 & 0.0302 \\ \hline
0.0 & 0.5 & 4 & 0.5 & -5.792 & -5.802 & 0.185 \\
0.0 & 0.5 & 4 & 0.35 & -5.885 & -5.891 & 0.105 \\
0.0 & 0.5 & 10 & 0.5 & -17.133 & -17.25 & 0.68 \\
0.0 & 0.5 & 10 & 0.35 & -18.849 & -18.934 & 0.448 \\ \hline
0.0 & 10.0 & 4 & 0.5 & -1.748 & -1.75 & 0.161 \\
0.0 & 10.0 & 4 & 0.35 & -1.829 & -1.829 & 0.0184 \\
0.0 & 10.0 & 10 & 0.5 & -8.218 & -8.341 & 1.48 \\
0.0 & 10.0 & 10 & 0.35 & -9.98 & -10.03 & 0.496 \\ \hline
0.0 & 20.0 & 4 & 0.5 & -1.492 & -1.739 & 14.2 \\
0.0 & 20.0 & 4 & 0.35 & -1.653 & -1.806 & 8.46 \\
0.0 & 20.0 & 10 & 0.5 & -8.202 & -8.328 & 1.51 \\
0.0 & 20.0 & 10 & 0.35 & -9.955 & -10.006 & 0.509 \\ \hline \hline
1.4 & 0.1 & 4 & 0.5 & -28.021 & -28.023 & 0.00697 \\
1.4 & 0.1 & 4 & 0.35 & -27.989 & -27.991 & 0.00755 \\
1.4 & 0.1 & 10 & 0.5 & -70.842 & -70.874 & 0.0453 \\
1.4 & 0.1 & 10 & 0.35 & -71.742 & -71.763 & 0.0291 \\ \hline
1.4 & 0.5 & 4 & 0.5 & -6.784 & -6.789 & 0.0609 \\
1.4 & 0.5 & 4 & 0.35 & -6.644 & -6.647 & 0.0327 \\
1.4 & 0.5 & 10 & 0.5 & -17.989 & -18.104 & 0.636 \\
1.4 & 0.5 & 10 & 0.35 & -19.445 & -19.534 & 0.456 \\ \hline
1.4 & 10.0 & 4 & 0.5 & -3.708 & -3.71 & 0.0728 \\
1.4 & 10.0 & 4 & 0.35 & -3.63 & -3.669 & 1.07 \\
1.4 & 10.0 & 10 & 0.5 & -10.067 & -10.243 & 1.71 \\
1.4 & 10.0 & 10 & 0.35 & -11.763 & -11.862 & 0.837 \\ \hline
1.4 & 20.0 & 4 & 0.5 & -3.673 & -3.681 & 0.218 \\
1.4 & 20.0 & 4 & 0.35 & -3.621 & -3.669 & 1.31 \\
1.4 & 20.0 & 10 & 0.5 & -10.028 & -10.224 & 1.92 \\
1.4 & 20.0 & 10 & 0.35 & -11.699 & -11.862 & 1.37 \\
\end{tabular}
\caption{Variational free energy defined in \eqref{eq:shift_var_energy} in the end of the training at epoch 500. 
Diff is $ (\mathcal{L}/(-\ln Z)) \times 100$, deviation from the exact free energy.
$\mathcal L$ is zero iff the variational approximation is exact.
%At $g/T = 0, \mu/g=0$, $N_x$ = 4 has significantly large deviation of the variational free energy. 
% $\hat{\mathcal{L}}  = \mathcal{L} - \ln Z$ is shifted loss function.
\label{tab:variational_free_energy}
}
\end{table}


In the later analysis, we do not include systematic error from the variation.
%This brings systematic error to the final results, which is propagated non-linearly and 
%it might affects. %Fortunately, we can compere our results for $\mu/g=0$ to the exact results [Sachs]. 

\subsection{Evaluation of large volume limit, and continuum extrapolation}
%We take the large volume limit.
In the previous work \cite{Chakraborty:2020uhf}, 
we observed that data from $N_x>10$ are necessary to take stable large volume limit.
However in this work, since we use state vector calculations for each variation step,  and we cannot obtain these data in practice.
We just use data from $N_x = 8, 10$ and evaluate the large volume limit.
%This will be improved by use of quantum device.

%\begin{figure}[h]
%\begin{center}
%\includegraphics[width=0.4\textwidth]{example-image-a}
%\end{center}
%\caption{
%Large volume limit. Statistical error from the sampling is shown.
%\label{fig:volume_limt}}
%\end{figure}

%Since data have large volume dependence, we do not take linear extrapolation for the volume limit.

\begin{figure}[h]
\begin{center}
\includegraphics[width=0.4\textwidth]{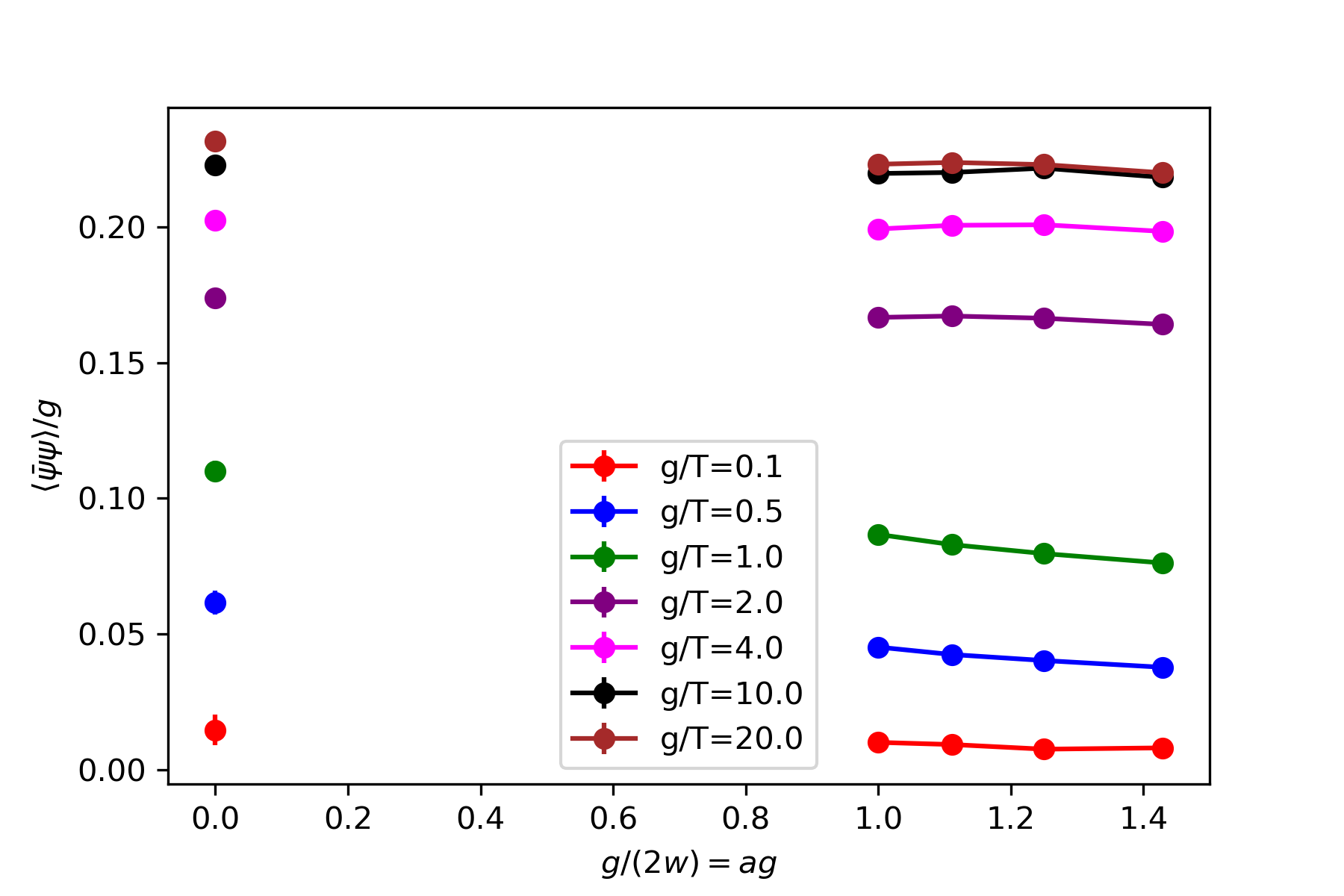}
\end{center}
\caption{
Continuum limit for $N_x=8$ and $\mu/g=0.0$. 
Different color corresponds to different temperature.
The error bar only contains statistical error.
\label{fig:continuum_limt_Nx8_mu0}}
\end{figure}

\begin{figure}[h]
\begin{center}
\includegraphics[width=0.4\textwidth]{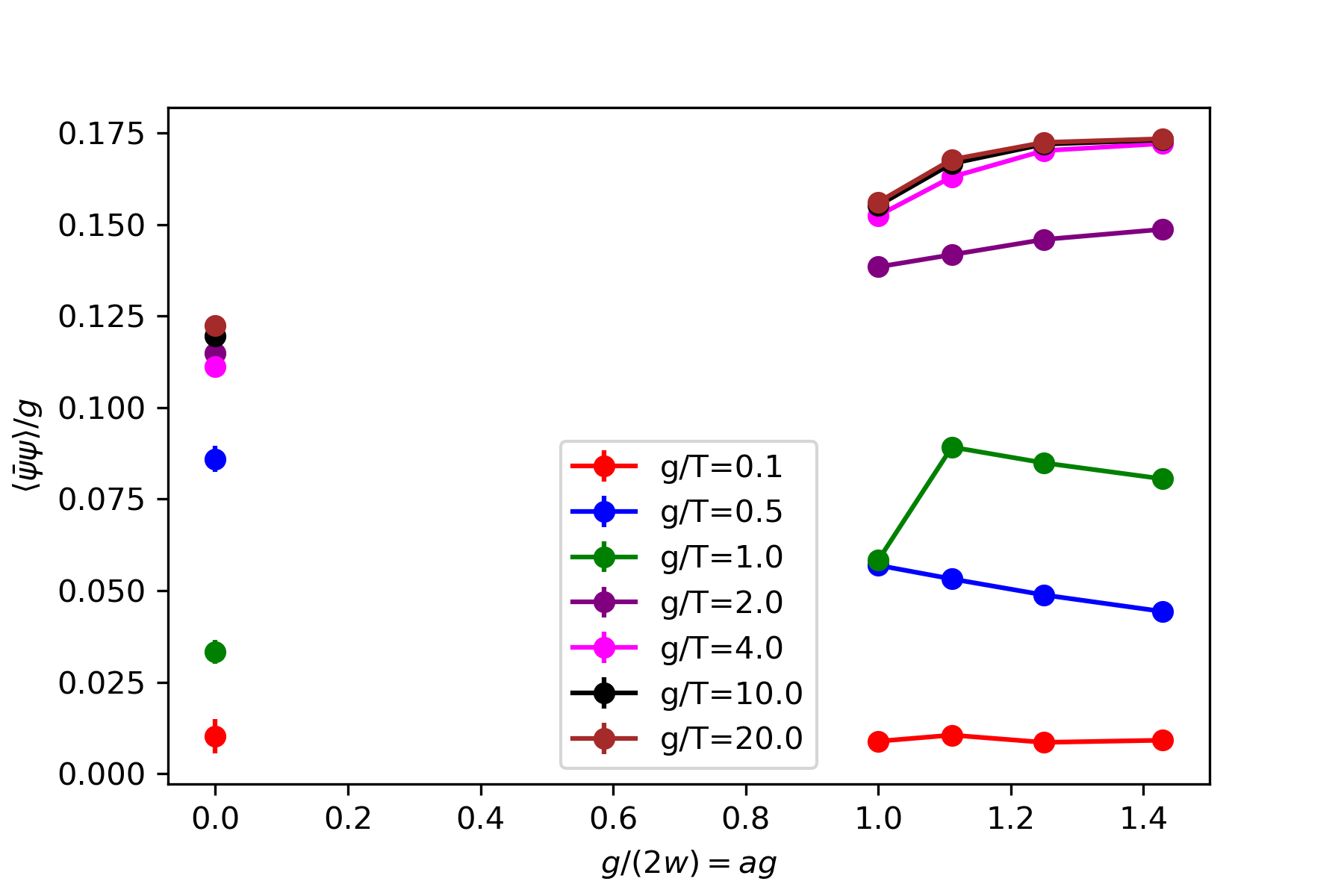}
\end{center}
\caption{
Same figure with Fig \ref{fig:continuum_limt_Nx8_mu0} but for $\mu/g = 1.4$.
\label{fig:continuum_limt_Nx8_mu1.4}}
\end{figure}

\begin{figure}[h]
\begin{center}
\includegraphics[width=0.4\textwidth]{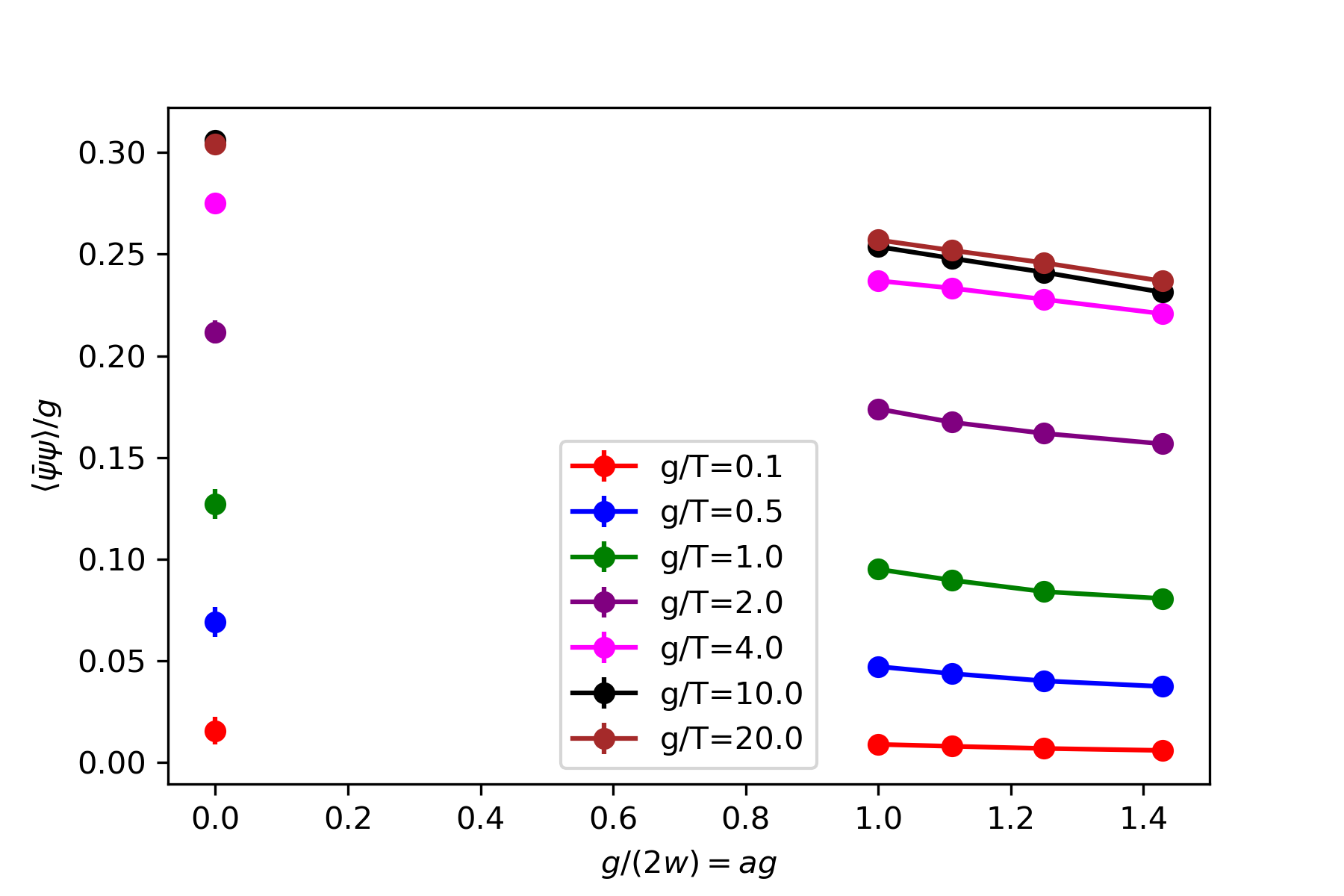}
\end{center}
\caption{
Same figure with Fig \ref{fig:continuum_limt_Nx8_mu0} but for $N_x = 10$.
\label{fig:continuum_limt_Nx10_mu0}}
\end{figure}

\begin{figure}[h]
\begin{center}
\includegraphics[width=0.4\textwidth]{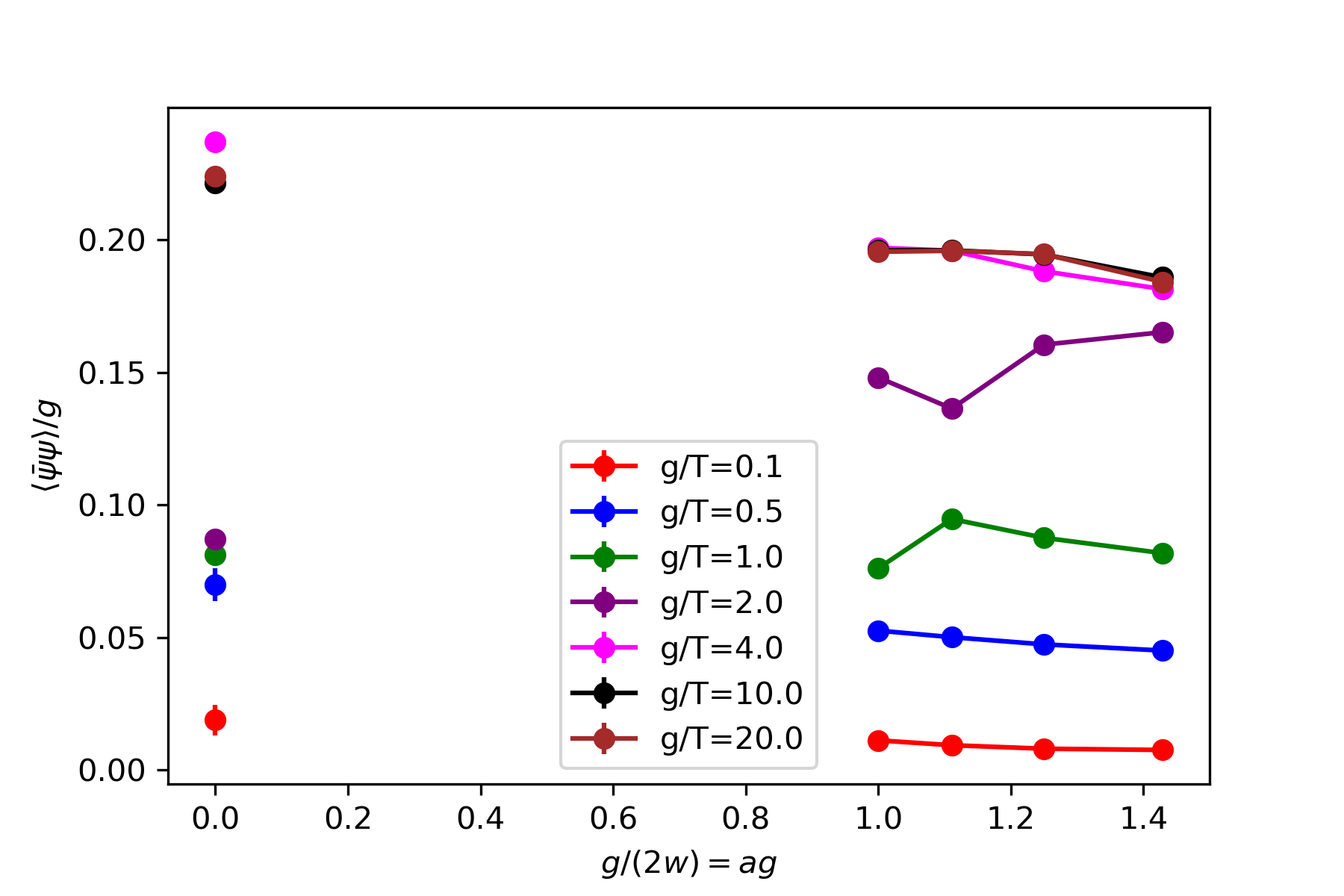}
\end{center}
\caption{
Same figure with Fig \ref{fig:continuum_limt_Nx8_mu0} but for $N_x = 10$ and $\mu/g = 1.4$.
\label{fig:continuum_limt_Nx10_mu1.4}}
\end{figure}

%\subsection{Continuum extrapolation}
%One can see from 
We extrapolate the data to $a\to0$ limit using linear fit ansatz and results are shown in 
Fig. \ref{fig:continuum_limt_Nx8_mu0} and Fig. \ref{fig:continuum_limt_Nx8_mu1.4} for $N_x = 8$ for $\mu/g = 0$ and $\mu/g=1.4$.
For $N_x = 10$, we plot in Fig. \ref{fig:continuum_limt_Nx10_mu0} and Fig. \ref{fig:continuum_limt_Nx10_mu1.4}.
The tendency of data is nearly linear, thus, this limit captures scaling to the continuum physics.


%- - - - -
\subsection{Phase diagram for $T$ -- $\mu$}
\begin{figure}[h]
\begin{center}
\includegraphics[width=0.5\textwidth]{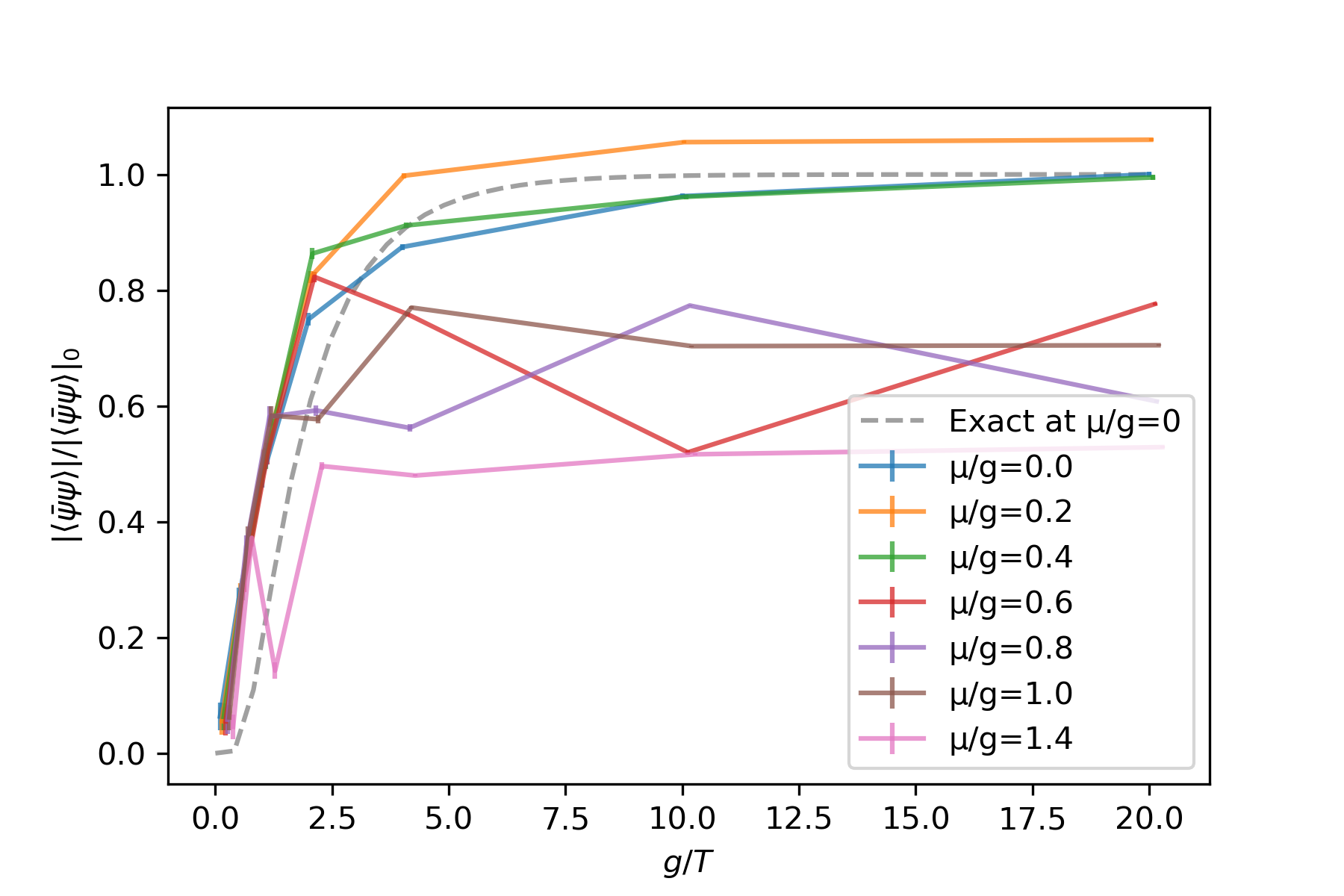}
\end{center}
\caption{
Chiral condensate as a function of $\beta$ at $a\to0$ for $N_x = 8$.
To avoid overlap, we shift the horizontal axis.
Error bar represents statistical error (same size with the symbols) but error from the variation is not included.
The results for the chiral condensate are normalized with the value in the continuum at $m=0$ with $g/T=20.0$, $\mu/g=0.0$ at $N_x=8$.
\label{fig:PbP_beta_nx8}}
\end{figure}

\begin{figure}[h]
\begin{center}
\includegraphics[width=0.5\textwidth]{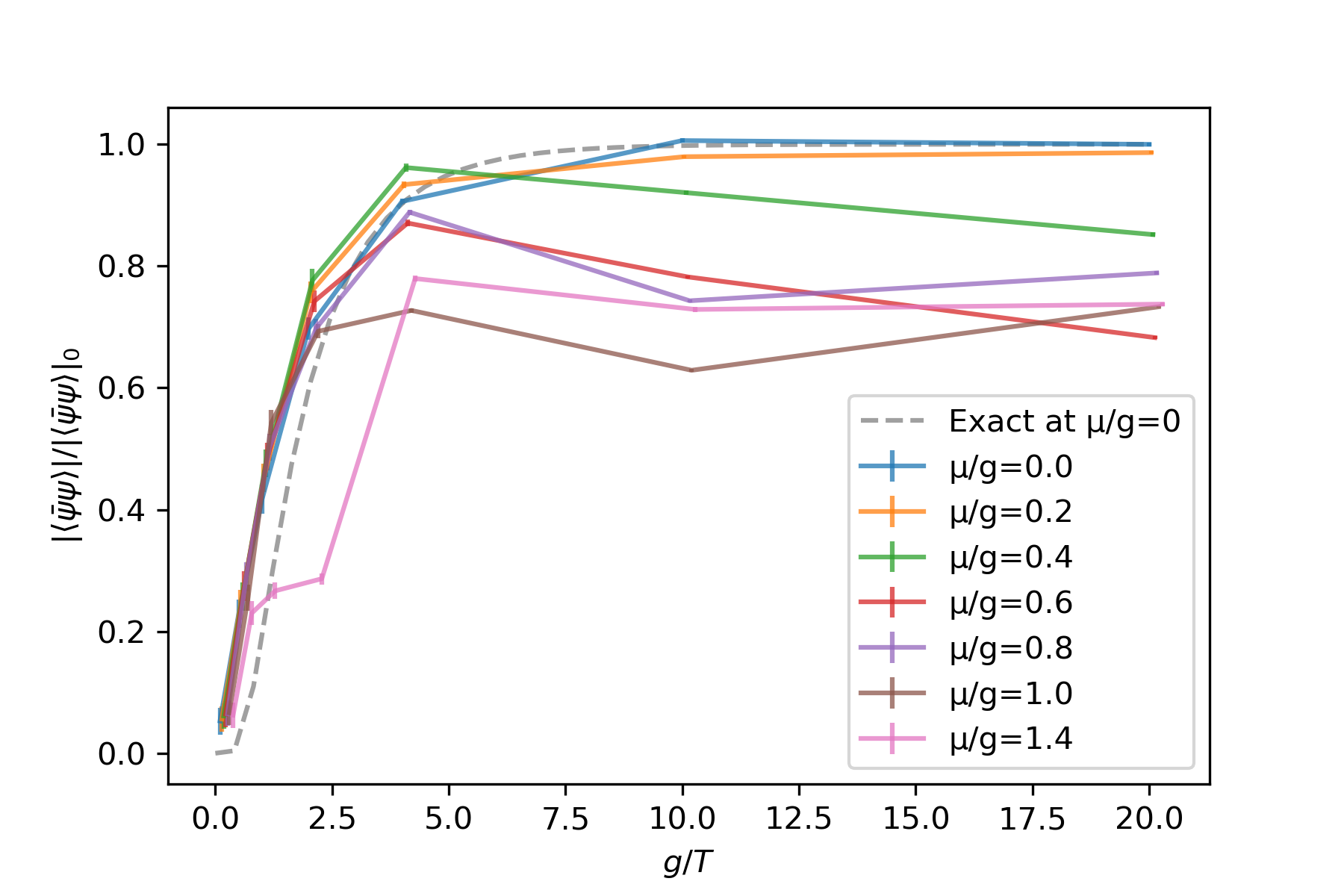}
\end{center}
\caption{
Same figure of Fig. \ref {fig:PbP_beta_nx8} but for $N_x = 10$.
\label{fig:PbP_beta_nx10}}
\end{figure}

Here we discuss phase diagram of the Schwinger model through  results for the chiral condensates.
We summarize the chiral condensate as a function of $\beta$ at $a\to0$ in 
Fig. \ref{fig:PbP_beta_nx8} and Fig. \ref{fig:PbP_beta_nx10} for $N_x = 8$ and 10, respectively.
To avoid overlap, we shift the horizontal axis and the vertical axis is normalized with the value in the continuum at $m=0$ with $g/T=20.0$, $\mu/g=0.0$ for each $N_x$ for our data.
Error bar represents statistical error but error from the variation is not included.
The error from the variation is estimated as $O(1)$ \% in the original data, so if we improve the variation,  we expect it will not be changed very much.
Our results for $\mu/g = 0$ is qualitatively agree with the exact results \cite{Sachs:1991en} in whole temperature range.
One can observe that, the chiral condensate gradually decreases along with $\mu/g$ for low temperature regime.
This has to be confirmed with larger lattice.

Density plot of the chiral condensate from \BetaVQE along with $T/g$ -- $\mu/g$ is shown in Fig. \ref{fig:phase-diagram}.
The central values in Fig. \ref{fig:PbP_beta_nx10} are interpolated and plotted.
The results for the chiral condensate are normalized with the value in the continuum at $m=0$ with $g/T=20.0$, $\mu/g=0.0$ for each $N_x$.
The massless Schwinger model does not have distinct phases but we phrase a parameter regime with the normalized chiral condensate larger than 1/2, as broken phase otherwise symmetric phase for showing purpose in the plot.
One can see that, this qualitatively agree with the what suggested for QCD \cite{CABIBBO197567, Fukushima:2010bq} for small chemical potential regime.


\section{Summary and discussion}
In this work, we investigate phase diagram on  finite temperature and density for the Schwinger model using a quantum classical hybrid algorithm called \BetaVQE, which is not affected by the sign problem.
Quantum expectation values are evaluated through 
state vector calculations on a classical computer which should be replaced by quantum calculations in the future.
Thanks to the state vector calculations, we evaluate the exact free energy $-\ln Z$ and we confirm that the variational algorithm gives $O(1) \%$ correct results.

%Continuum limit are evaluated but only constant fits have been taken because of stong scaling dependence.
Only continuum limit is taken and large volume limit has not been taken in this work.
%This is probably because lack of data from large volume \cite{Chakraborty:2020uhf}. 
While, we observe qualitative agreement along with the temperature to the exact results for $\mu/g=0$ \cite{Sachs:1991en}
and the deviation is similar to \cite{Banuls:2016lkq}.
To establish physics at infinity volume, we should replace the state-vector calculations to a tensor network or quantum device to overcome the shortage of the numerical resource.
We enphasize that \BetaVQE can be used with a quantum device, and once we get a trustable quantum machine, fault-tolerant quantum computer, we can easily apply this calculation on it.

We have not investigated crystal structure for $\mu/g > 0$ which is predicted in several studies \cite{Metlitski:2006id, Maedan:2009yi, Narayanan:2012qf}.
By defining position dependent gauge invariant correlators, and use a large number of sampling, we could see the crystal structure and dependence on the temperature.

We have used the autoregressive model \cite{autogressive2015} to realize the probability distribution $p_\phi (x)$ in the density matrix.
It could be replaced by other neural net model like the normalizing flow \cite{Albergo:2019eim}.
Use of QITE \cite{motta2020determining} is also interested.

\section*{Acknowledgement}
AT thanks to Lei Wang and Jinguo Liu for helping me developing the code.
Numerical computation in this work was carried out at the Yukawa Institute Computer Facility.
The work of AT was supported by  JSPS KAKENHI Grant No. JP20K14479.

\bibliography{ref}

\begin{thebibliography}{10}

\bibitem{CABIBBO197567}
N.~Cabibbo and G.~Parisi.
\newblock Exponential hadronic spectrum and quark liberation.
\newblock {\em Physics Letters B}, 59(1):67--69, 1975.

\bibitem{Fukushima:2010bq}
Kenji Fukushima and Tetsuo Hatsuda.
\newblock {The phase diagram of dense QCD}.
\newblock {\em Rept. Prog. Phys.}, 74:014001, 2011.

\bibitem{Guenther:2020jwe}
Jana~N. Guenther.
\newblock {Overview of the QCD phase diagram: Recent progress from the
  lattice}.
\newblock {\em Eur. Phys. J. A}, 57(4):136, 2021.

\bibitem{HotQCD:2012fhj}
A.~Bazavov et~al.
\newblock {Fluctuations and Correlations of net baryon number, electric charge,
  and strangeness: A comparison of lattice QCD results with the hadron
  resonance gas model}.
\newblock {\em Phys. Rev. D}, 86:034509, 2012.

\bibitem{Ding:2015ona}
Heng-Tong Ding, Frithjof Karsch, and Swagato Mukherjee.
\newblock {Thermodynamics of strong-interaction matter from Lattice QCD}.
\newblock {\em Int. J. Mod. Phys. E}, 24(10):1530007, 2015.

\bibitem{HotQCD:2018pds}
A.~Bazavov et~al.
\newblock {Chiral crossover in QCD at zero and non-zero chemical potentials}.
\newblock {\em Phys. Lett. B}, 795:15--21, 2019.

\bibitem{Borsanyi:2018grb}
Szabolcs Borsanyi, Zoltan Fodor, Jana~N. Guenther, Sandor~K. Katz, Kalman~K.
  Szabo, Attila Pasztor, Israel Portillo, and Claudia Ratti.
\newblock {Higher order fluctuations and correlations of conserved charges from
  lattice QCD}.
\newblock {\em JHEP}, 10:205, 2018.

\bibitem{Karsch:2019mbv}
Frithjof Karsch.
\newblock {Critical behavior and net-charge fluctuations from lattice QCD}.
\newblock {\em PoS}, CORFU2018:163, 2019.

\bibitem{Goswami:2020yez}
Jishnu Goswami, Frithjof Karsch, Christian Schmidt, Swagato Mukherjee, and
  Peter Petreczky.
\newblock {Comparing conserved charge fluctuations from lattice QCD to HRG
  model calculations}.
\newblock {\em Acta Phys. Polon. Supp.}, 14:251, 2021.

\bibitem{Gattringer:2015nea}
Christof Gattringer, Thomas Kloiber, and Vasily Sazonov.
\newblock {Solving the sign problems of the massless lattice Schwinger model
  with a dual formulation}.
\newblock {\em Nucl. Phys. B}, 897:732--748, 2015.

\bibitem{Ratti:2019tvj}
Claudia Ratti.
\newblock {QCD at non-zero density and phenomenology}.
\newblock {\em PoS}, LATTICE2018:004, 2019.

\bibitem{Alexandru:2020wrj}
Andrei Alexandru, Gokce Basar, Paulo~F. Bedaque, and Neill~C. Warrington.
\newblock {Complex paths around the sign problem}.
\newblock {\em Rev. Mod. Phys.}, 94(1):015006, 2022.

\bibitem{Aarts:2017vrv}
Gert Aarts, Erhard Seiler, Denes Sexty, and Ion-Olimpiu Stamatescu.
\newblock {Complex Langevin dynamics and zeroes of the fermion determinant}.
\newblock {\em JHEP}, 05:044, 2017.
\newblock [Erratum: JHEP 01, 128 (2018)].

\bibitem{Seiler:2017wvd}
Erhard Seiler.
\newblock {Status of Complex Langevin}.
\newblock {\em EPJ Web Conf.}, 175:01019, 2018.

\bibitem{Berger:2019odf}
Casey~E. Berger, Lukas Rammelm\"uller, Andrew~C. Loheac, Florian Ehmann, Jens
  Braun, and Joaqu\'\i{}n~E. Drut.
\newblock {Complex Langevin and other approaches to the sign problem in quantum
  many-body physics}.
\newblock {\em Phys. Rept.}, 892:1--54, 2021.

\bibitem{Scherzer:2020kiu}
M.~Scherzer, D.~Sexty, and I.~O. Stamatescu.
\newblock {Deconfinement transition line with the complex Langevin equation up
  to $\mu /T \sim 5$}.
\newblock {\em Phys. Rev. D}, 102(1):014515, 2020.

\bibitem{Witten:2010cx}
Edward Witten.
\newblock {Analytic Continuation Of Chern-Simons Theory}.
\newblock {\em AMS/IP Stud. Adv. Math.}, 50:347--446, 2011.

\bibitem{Mori:2017zyl}
Yuto Mori, Kouji Kashiwa, and Akira Ohnishi.
\newblock {Lefschetz thimbles in fermionic effective models with repulsive
  vector-field}.
\newblock {\em Phys. Lett. B}, 781:688--693, 2018.

\bibitem{Schmidt:2017gvu}
Christian Schmidt and Felix Ziesch\'e.
\newblock {Simulating low dimensional QCD with Lefschetz thimbles}.
\newblock {\em PoS}, LATTICE2016:076, 2017.

\bibitem{Fujisawa:2021hxh}
Genki Fujisawa, Jun Nishimura, Katsuta Sakai, and Atis Yosprakob.
\newblock {Backpropagating Hybrid Monte Carlo algorithm for fast Lefschetz
  thimble calculations}.
\newblock {\em JHEP}, 04:179, 2022.

\bibitem{Fukuma:2022yhy}
Masafumi Fukuma, Nobuyuki Matsumoto, and Yusuke Namekawa.
\newblock Numerical sign problem and the tempered lefschetz thimble method.
\newblock In {\em 21st Hellenic School and Workshops on Elementary Particle
  Physics and Gravity}, 5 2022.

\bibitem{Mori:2017nwj}
Yuto Mori, Kouji Kashiwa, and Akira Ohnishi.
\newblock {Application of a neural network to the sign problem via the path
  optimization method}.
\newblock {\em PTEP}, 2018(2):023B04, 2018.

\bibitem{Alexandru:2017czx}
Andrei Alexandru, Paulo~F. Bedaque, Henry Lamm, and Scott Lawrence.
\newblock {Deep Learning Beyond Lefschetz Thimbles}.
\newblock {\em Phys. Rev. D}, 96(9):094505, 2017.

\bibitem{Kashiwa:2018vxr}
Kouji Kashiwa, Yuto Mori, and Akira Ohnishi.
\newblock {Controlling the model sign problem via the path optimization method:
  Monte Carlo approach to a QCD effective model with Polyakov loop}.
\newblock {\em Phys. Rev. D}, 99(1):014033, 2019.

\bibitem{Lawrence:2021izu}
Scott Lawrence and Yukari Yamauchi.
\newblock {Normalizing Flows and the Real-Time Sign Problem}.
\newblock {\em Phys. Rev. D}, 103(11):114509, 2021.

\bibitem{Detmold:2021ulb}
William Detmold, Gurtej Kanwar, Henry Lamm, Michael~L. Wagman, and Neill~C.
  Warrington.
\newblock {Path integral contour deformations for observables in $SU(N)$ gauge
  theory}.
\newblock {\em Phys. Rev. D}, 103(9):094517, 2021.

\bibitem{Kanwar:2021tkd}
Gurtej Kanwar and Michael~L. Wagman.
\newblock {Real-time lattice gauge theory actions: Unitarity, convergence, and
  path integral contour deformations}.
\newblock {\em Phys. Rev. D}, 104(1):014513, 2021.

\bibitem{Namekawa:2021nzu}
Yusuke Namekawa, Kouji Kashiwa, Akira Ohnishi, and Hayato Takase.
\newblock {Gauge invariant input to neural network for path optimization
  method}.
\newblock {\em Phys. Rev. D}, 105(3):034502, 2022.

\bibitem{Lawrence:2020irw}
Scott Lawrence.
\newblock {\em Sign Problems in Quantum Field Theory: Classical and Quantum
  Approaches}.
\newblock PhD thesis, Maryland U., 2020.

\bibitem{Martinez:2016yna}
E.~A. Martinez et~al.
\newblock {Real-time dynamics of lattice gauge theories with a few-qubit
  quantum computer}.
\newblock {\em Nature}, 534:516--519, 2016.

\bibitem{Huffman:2021neh}
Emilie Huffman, Miguel~Garc\'\i{}a Vera, and Debasish Banerjee.
\newblock {Real-time dynamics of Plaquette Models using NISQ Hardware}.
\newblock 9 2021.

\bibitem{Chakraborty:2020uhf}
Bipasha Chakraborty, Masazumi Honda, Taku Izubuchi, Yuta Kikuchi, and Akio
  Tomiya.
\newblock {Classically emulated digital quantum simulation of the Schwinger
  model with a topological term via adiabatic state preparation}.
\newblock {\em Phys. Rev. D}, 105(9):094503, 2022.

\bibitem{Bharti:2021zez}
Kishor Bharti et~al.
\newblock {Noisy intermediate-scale quantum algorithms}.
\newblock {\em Rev. Mod. Phys.}, 94(1):015004, 2022.

\bibitem{Yamamoto:2021vxp}
Arata Yamamoto.
\newblock {Quantum variational approach to lattice gauge theory at nonzero
  density}.
\newblock {\em Phys. Rev. D}, 104(1):014506, 2021.

\bibitem{motta2020determining}
Mario Motta, Chong Sun, Adrian~TK Tan, Matthew~J O’Rourke, Erika Ye, Austin~J
  Minnich, Fernando~GSL Brandao, and Garnet~Kin Chan.
\newblock Determining eigenstates and thermal states on a quantum computer
  using quantum imaginary time evolution.
\newblock {\em Nature Physics}, 16(2):205--210, 2020.

\bibitem{Liu_2019}
Jin-Guo Liu, Yi-Hong Zhang, Yuan Wan, and Lei Wang.
\newblock Variational quantum eigensolver with fewer qubits.
\newblock {\em Physical Review Research}, 1(2), sep 2019.

\bibitem{Ciavarella:2021lel}
Anthony~N. Ciavarella and Ivan~A. Chernyshev.
\newblock {Preparation of the SU(3) lattice Yang-Mills vacuum with variational
  quantum methods}.
\newblock {\em Phys. Rev. D}, 105(7):074504, 2022.

\bibitem{Deutsch_2018}
Joshua~M Deutsch.
\newblock Eigenstate thermalization hypothesis.
\newblock {\em Reports on Progress in Physics}, 81(8):082001, jul 2018.

\bibitem{Sugiura:2011hm}
Sho Sugiura and Akira Shimizu.
\newblock {Thermal Pure Quantum States at Finite Temperature}.
\newblock {\em Phys. Rev. Lett.}, 108:240401, 2012.

\bibitem{liu2021solving}
Jin-Guo Liu, Liang Mao, Pan Zhang, and Lei Wang.
\newblock Solving quantum statistical mechanics with variational autoregressive
  networks and quantum circuits.
\newblock {\em Machine Learning: Science and Technology}, 2(2):025011, 2021.

\bibitem{Schwinger:1962tp}
Julian~S. Schwinger.
\newblock {Gauge Invariance and Mass. 2.}
\newblock {\em Phys. Rev.}, 128:2425--2429, 1962.

\bibitem{Pak:1977an}
Namik~K. Pak and P.~Senjanovic.
\newblock {Axial Anomaly and Chiral Symmetry Breaking in Schwinger Model: A
  Canonical Approach}.
\newblock {\em Phys. Lett. B}, 71:333--336, 1977.

\bibitem{Johnson:1963vz}
K.~Johnson.
\newblock {gamma(5) invariance}.
\newblock {\em Phys. Lett.}, 5:253--255, 1963.

\bibitem{Sachs:1991en}
Ivo Sachs and Andreas Wipf.
\newblock {Finite temperature Schwinger model}.
\newblock {\em Helv. Phys. Acta}, 65:652--678, 1992.

\bibitem{Finkenrath:2022ogg}
Jacob Finkenrath.
\newblock {Tackling critical slowing down using global correction steps with
  equivariant flows: the case of the Schwinger model}.
\newblock 1 2022.

\bibitem{Albergo:2022qfi}
Michael~S. Albergo, Denis Boyda, Kyle Cranmer, Daniel~C. Hackett, Gurtej
  Kanwar, S\'ebastien Racani\`ere, Danilo~J. Rezende, Fernando Romero-L\'opez,
  Phiala~E. Shanahan, and Julian~M. Urban.
\newblock Flow-based sampling in the lattice schwinger model at criticality.
\newblock 2 2022.

\bibitem{Banuls:2013jaa}
M.~C. Ba\~nuls, K.~Cichy, K.~Jansen, and J.~I. Cirac.
\newblock {The mass spectrum of the Schwinger model with Matrix Product
  States}.
\newblock {\em JHEP}, 11:158, 2013.

\bibitem{Banuls:2016lkq}
Mari~Carmen Ba\~nuls, Krzysztof Cichy, Karl Jansen, and Hana Saito.
\newblock {Chiral condensate in the Schwinger model with Matrix Product
  Operators}.
\newblock {\em Phys. Rev. D}, 93(9):094512, 2016.

\bibitem{Funcke:2019zna}
Lena Funcke, Karl Jansen, and Stefan K\"uhn.
\newblock {Topological vacuum structure of the Schwinger model with matrix
  product states}.
\newblock {\em Phys. Rev. D}, 101(5):054507, 2020.

\bibitem{Butt:2019uul}
Nouman Butt, Simon Catterall, Yannick Meurice, Ryo Sakai, and Judah
  Unmuth-Yockey.
\newblock Tensor network formulation of the massless schwinger model with
  staggered fermions.
\newblock {\em Phys. Rev. D}, 101(9):094509, 2020.

\bibitem{Fukuma:2021cni}
Masafumi Fukuma, Daisuke Kadoh, and Nobuyuki Matsumoto.
\newblock Tensor network approach to two-dimensional yang-mills theories.
\newblock {\em PTEP}, 2021(12):123B03, 2021.

\bibitem{Shaw:2020udc}
Alexander~F. Shaw, Pavel Lougovski, Jesse~R. Stryker, and Nathan Wiebe.
\newblock {Quantum Algorithms for Simulating the Lattice Schwinger Model}.
\newblock {\em Quantum}, 4:306, 2020.

\bibitem{Banuls:2016gid}
Mari~Carmen Ba\~nuls, Krzysztof Cichy, J.~Ignacio Cirac, Karl Jansen, and
  Stefan K\"uhn.
\newblock {Density Induced Phase Transitions in the Schwinger Model: A Study
  with Matrix Product States}.
\newblock {\em Phys. Rev. Lett.}, 118(7):071601, 2017.

\bibitem{autogressive2015}
Mathieu Germain, Karol Gregor, Iain Murray, and Hugo Larochelle.
\newblock {MADE:} masked autoencoder for distribution estimation.
\newblock {\em CoRR}, abs/1502.03509, 2015.

\bibitem{Kanwar:2020xzo}
Gurtej Kanwar, Michael~S. Albergo, Denis Boyda, Kyle Cranmer, Daniel~C.
  Hackett, S\'ebastien Racani\`ere, Danilo~Jimenez Rezende, and Phiala~E.
  Shanahan.
\newblock {Equivariant flow-based sampling for lattice gauge theory}.
\newblock {\em Phys. Rev. Lett.}, 125(12):121601, 2020.

\bibitem{Albergo:2019eim}
M.~S. Albergo, G.~Kanwar, and P.~E. Shanahan.
\newblock {Flow-based generative models for Markov chain Monte Carlo in lattice
  field theory}.
\newblock {\em Phys. Rev. D}, 100(3):034515, 2019.

\bibitem{Boyda:2020hsi}
Denis Boyda, Gurtej Kanwar, S\'ebastien Racani\`ere, Danilo~Jimenez Rezende,
  Michael~S. Albergo, Kyle Cranmer, Daniel~C. Hackett, and Phiala~E. Shanahan.
\newblock {Sampling using $SU(N)$ gauge equivariant flows}.
\newblock {\em Phys. Rev. D}, 103(7):074504, 2021.

\bibitem{adamopt}
Diederik~P. Kingma and Jimmy Ba.
\newblock Adam: A method for stochastic optimization, 2014.

\bibitem{shiftrule1}
Gavin~E Crooks.
\newblock Gradients of parameterized quantum gates using the parameter-shift
  rule and gate decomposition, 2019.

\bibitem{Banchi_2021}
Leonardo Banchi and Gavin~E. Crooks.
\newblock Measuring analytic gradients of general quantum evolution with the
  stochastic parameter shift rule.
\newblock {\em Quantum}, 5:386, jan 2021.

\bibitem{Metlitski:2006id}
Max~A. Metlitski.
\newblock {Is Schwinger model at finite density a crystal?}
\newblock {\em Phys. Rev. D}, 75:045004, 2007.

\bibitem{Maedan:2009yi}
Shinji Maedan.
\newblock {Influence of current mass on the spatially inhomogeneous chiral
  condensate}.
\newblock {\em Prog. Theor. Phys.}, 123:285--302, 2010.

\bibitem{Narayanan:2012qf}
R.~Narayanan.
\newblock {Two flavor massless Schwinger model on a torus at a finite chemical
  potential}.
\newblock {\em Phys. Rev. D}, 86:125008, 2012.

\bibitem{betaVQEjl}
{BetaVQE.jl}.
\newblock \url{https://github.com/wangleiphy/BetaVQE.jl}.

\bibitem{Luo2020yaojlextensible}
Xiu-Zhe Luo, Jin-Guo Liu, Pan Zhang, and Lei Wang.
\newblock Yao.jl: {E}xtensible, {E}fficient {F}ramework for {Q}uantum
  {A}lgorithm {D}esign.
\newblock {\em {Quantum}}, 4:341, October 2020.

\bibitem{Zygotejl}
Mike Innes, Alan Edelman, Keno Fischer, Chris Rackauckas, Elliot Saba, Viral~B
  Shah, and Will Tebbutt.
\newblock A differentiable programming system to bridge machine learning and
  scientific computing, 2019.

\bibitem{julialang}
Jeff Bezanson, Alan Edelman, Stefan Karpinski, and Viral~B. Shah.
\newblock Julia: A fresh approach to numerical computing.
\newblock {\em SIAM Review}, 59(1):65--98, 2017.

\end{thebibliography}

\end{document}